\documentclass[12pt]{article}
\usepackage{graphicx}
\usepackage{amsmath}

\title{ KALMAN FILTER BASED TRACKER STUDY FOR LEPTON FLAVOR VIOLATION EXPERIMENTS }

\author{\textbf{Rashid M. Djilkibaev$^{1}$ \thanks{Permanent address:
     Institute for Nuclear Research, 60-th Oct. pr. 7a,
     Moscow 117312, Russia}\ ,
     Rostislav V. Konoplich$^{1,2}$}\\
   \normalsize$^{1}$Department of Physics, New York University,
   New York, NY 10003\\
   \normalsize$^{2}$Manhattan College, Riverdale, New York, NY, 10471}

\begin{document}

\maketitle

\begin{abstract}

A tracking detector is proposed for
lepton flavor violation experiments 
($\mu \to e$ conversion, $\mu \to e + \gamma$, $\mu \to 3e $)
consisting  of identical chambers which can be reconfigured to
meet the  requirements for all three experiments.
A pattern recognition and  track reconstruction
procedure  based on the Kalman filter technique
is presented for this detector.

The pattern recognition proceeds in two stages.
At the first stage only  
hit straw tube center coordinates, without drift time information, are
used to reduce the background to a manageable level. 
At the second stage the drift time information is incorporated and a deterministic annealing filter
is applied to reach the final level of background suppression.
The final track momentum reconstruction is provided by a combinatorial drop filter
which is effective in hit-to-track assignment.

The momentum resolution of the tracker in measuring monochromatic leptons 
is found to be  $\sigma_{p}$ = 0.17  and 0.26 MeV for the $\mu \to e$ conversion
and $\mu^+ \to e^+ + \gamma$ processes, respectively.
The tracker  reconstruction resolution for the total scalar lepton momentum 
is  $\sigma_{p} = $ 0.33 MeV for the $\mu \to 3e$ process.
The obtained tracker resolutions allow an increase in sensitivity to the branching ratios 
for these processes by
a few orders of magnitude over current experimental limits.

\end{abstract}

\section{Introduction}

The goal of this work is to 
demonstrate a pattern recognition and  track reconstruction
procedure for a tracking detector which provides sufficient resolution 
to be used in  new lepton flavor violation experiments
sensitive to branching ratios 
 a few order of magnitude lower than current experimental limits.
The observation of one of the three processes
$\mu \to e$ conversion, $\mu \to e + \gamma$ and $\mu \to 3e $ 
would provide the first
direct evidence for lepton flavor violation in the charged lepton sector 
\cite{ver,kuno}
and require new physics,
beyond the Standard Model.
To reach such high sensitivity  a substantial improvement in muon beam intensity \cite{melc}
is required. This leads to the need for new designs of tracking detectors. 
A straw tube tracker proposed in this work consists of a set of identical  chambers. 
A simple reconfiguration 
of the chambers  allows   meeting requirements for all three lepton flavor violation  experiments.

The signature of the $\mu \to e$ conversion ($\mu^{-} + N \to e^{-} + N$) process is
clear: a single  monochromatic electron in the final state with  energy 
close to the muon mass $m_{\mu}$. The  $\mu^+ \to e^+ + \gamma$ process has in the final state 
a  monochromatic positron and a photon with
an energy equal to half the muon mass. 
It is assumed that an external trigger is provided for the tracker.
Photon reconstruction is not considered in this article 
because this study is devoted to pattern recognition and track reconstruction of
electrons and positrons.
The $\mu^+ \to e^+ + e^+ + e^-$ process has in the final state
an electron and two positrons with  energies from 0 to $m_{\mu}$/2 and  an average energy equal  
to one third of the muon mass.

In this work a pattern recognition and  tracker reconstruction study was done  
for the  $\mu \to e$ conversion process, in the presence of background.
A tracker resolution study for charged particles in $\mu^+ \to e^+ + \gamma$  and $\mu^+ \to e^+ + e^+ + e^-$ processes
was done without background.  
Our analysis is based on a full GEANT3 \cite{geant}  simulation taking into account 
 individual straw structure.

\section{Tracker description}

\begin{figure}[htb!]
  \centering
  \includegraphics[width=0.7\textwidth]{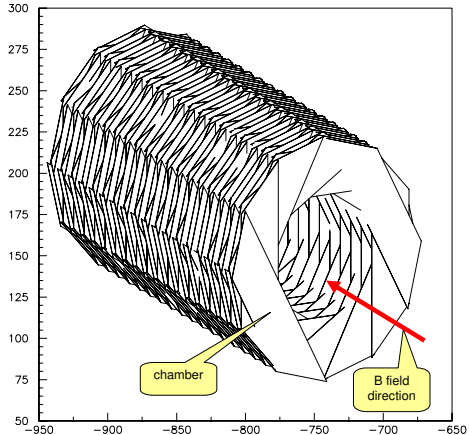} 
  \caption[Short caption.]{
  Schematic drawing of the tracker design.
}
\label{fig:setup}
\end{figure}

Muons stopping in a target produce electrons or positrons depending on the physical
process. These charged particles move in a uniform magnetic field following
helical trajectories.  
A tracking detector is used to measure the particle trajectory and thereby determine its momentum.
The proposed tracker (Fig. \ref{fig:setup}) consists of  chambers with straw tubes placed 
transverse to a uniform magnetic field,
which is along the tracker axis. 
Each straw tube
measures the drift time corresponding to the radial distance at the closest approach
of a charged particle from the sense wire.
The tracker is operated in vacuum.
The  high muon beam intensity forces a tracker design without matter in  a cylindrical central zone,
to let the beam pass through the tracker. 
To get more redundancy for very rare signal events it is required that the measured trajectory 
should have two full turns in the tracker.
This requirement sets   
limit on the minimal tracker length of about 300 cm.

A tracker plane consists of two  trapezoidal straw tube chambers (see Fig. \ref{fig:setup}).  
The planes are distributed uniformly over the length of the tracker.
Rotating consecutive planes by an  angle of 30$^{o}$ gives an effective ``stereo'' 
of crossed directions for 12 different views.
Each chamber consists of one layer of straw tubes (60 straws) of 5
mm diameter.
The straws are assumed to have wall 
thickness 25 $\mu$m and are
constructed of kapton.
Isobutane (C$_{4}$H$_{10}$) gas is assumed to fill the tubes. 
The tracker design allows changing the central zone radius, by moving the 
chambers closer to or further from the central axis, and changing the 
number of chambers along the tracker axis.
This is needed to meet the different requirements for 
registration of electrons or positrons in the final state 
of each lepton flavor violation experiment. 

Two tracker chamber layouts and magnetic field configurations are studied: 
one for the $\mu \to e$  conversion process and the second 
for $\mu \to e + \gamma$  and $\mu \to e + e + e$ processes. 
The first tracker layout contains 108 planes and has the central zone radius of 38 cm. 
The second tracker layout contains 54 planes and has the central zone radius of  10 cm. 
The tracker length for both layouts is the same, 300 cm.
In the two layouts the tracker is immersed 
in a magnetic field  of 1 T and 0.5 T, respectively. 

\section{Pattern recognition}

Track reconstruction in lepton flavor violation experiments faces a significant amount of
background hits in a tracker because of the high beam intensity needed to detect rare processes. 
A two stage procedure was developed to provide the pattern
recognition in the tracker and to suppress background hits. 
At the first  stage of pattern recognition only
information on the centers of hit straws is used.
The result of this stage is a significant suppression of 
background hits. Also, an approximate helix
 fitted to the straw hit centers is found and used in the next stage of pattern recognition.
To improve the suppression of background hits the second stage of 
pattern recognition uses drift time. 
A deterministic annealing 
filter DAF  \cite{daf, rev} is applied at the second stage.
The DAF effectively suppresses backgrounds by dealing simultaneously 
with multiple competing hits.

\subsection{Pattern recognition  without drift time information}

\begin{figure}[htb!]
  \centering
  \includegraphics[width=0.7\textwidth]{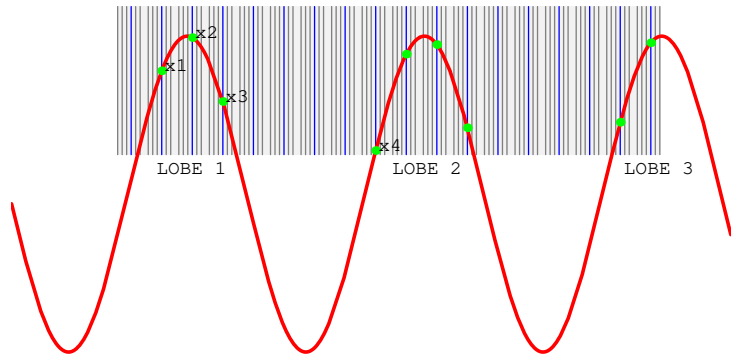} 
  \caption[Short caption.]{
  Lobes in the tracker view. Points show straw hit centers.
  The sinusoidal projection of a helix  is shown in red. 
  The starting point of the track is not shown in the figure.
}
\label{fig:sin1}
\end{figure}

A detailed pattern recognition and tracker reconstruction study was done for  monochromatic electrons of 105 MeV
from  the $\mu \to e$ conversion process
in the presence of background hits.
In a uniform magnetic field the trajectory of a charged particle is a helix described
by 5 parameters.
The construction of the tracker allows significant simplification of the initial
step of the pattern recognition by reducing the problem to two dimensional
tracker views. Twelve tracker views are formed by planes through the tracker axis
normal to the straws.
 In a typical
event, a two-dimensional projection of the helical trajectory, a
sine curve, is observed in a few views. 
There are
approximately 10 hits in each view, and 30 hits in the average in the
event.
The average number of background hits is 300 per event, which corresponds to an average
straw hit rate of about 800 kHz.
 The hits are grouped in lobes (see Fig. \ref{fig:sin1}) with a typical
gap between lobes of about 70 cm.
As seen in Fig. \ref{fig:sin1} the sensitive area of the tracker measures a small part of a track,
but  combining hits from two lobes significantly improves the precision of the momentum reconstruction.

The sinusoidal projection of a helix in each view is described by four parameters:
$x_{0}^{\prime }$ , $z_{0}^{\prime }$, $P_{L}$, $P_{T}$,

\begin{equation}
x_{i}^{\prime }=x_{0}^{\prime } + a P_{T}\cos (\frac{z_{i}-z_{0}^{\prime }}{a P_{L}})
 \label{eq0}
\end{equation}

where $P_{L}$ and $P_{T}$ are the longitudinal and transversal
momenta. The constant  a equals 1/(2.998 $B$), where $B$ is the magnetic field measured in Tesla.
 The coordinate $z$ is common to all projections. 
Parameters $x_{0}^{\prime}$ and $z_{0}^{\prime}$ 
are defined for each given view such that the particle is created at the target with
coordinate ($x_{0}^{\prime}+ a P_{T}$, $z_{0}^{\prime}$).

The reconstruction algorithm 
starts from a single tracker view. 
Four hits from this  view are used to give a set of equations
for helix parameters in the coordinate system related to that view.
The system of four equations is 
solved to get the parameters $P_{L},P_{T},z_{0}^{\prime},x_{0}^{\prime }$ \cite{hep-ex}.
All possible four hit combinations are considered.
Only the combinations that survive a cut-off in the
particle momentum 
($\pm 20 \% $ of the expected value)
are retained for subsequent analysis.

Adding a fifth hit from a different view than
the four hit combination  allows   defining a helical
3-D trajectory for the five-hit combination.

The following approach is used to select signal hits.
If we select a combination with five signal hits the defined helical trajectory should correlate in space 
with  other signal hits.
For further analysis a five-hit combination is selected if
the helical trajectory matches at least 15 additional hits in the road ($\pm$1 straw).
A hit is rejected if it is not in any selected five-hit combinations.
 Due to the strong spatial correlations between the
signal hits in comparison with the un-correlated background hits, the
number of background hits is reduced drastically by applying this
road requirement.
A fit is applied to reconstruct an average helical trajectory
on the basis of
the list of the selected tracker hits.
This fit does not take into account multiple scattering.

At the first stage of pattern recognition procedure
the overall background rejection factor is about 130.
The electron momentum from the $\mu \to e$ conversion process 
can be reconstructed    
in this stage with a standard deviation  
$\sigma_{p} $ = 0.45 MeV 
corresponding to  relative momentum resolution  $\sigma_{p} /p = 0.5\%$.

\begin{figure}[htb!]
  \centering{\hbox{
  \includegraphics[width=0.5\textwidth]{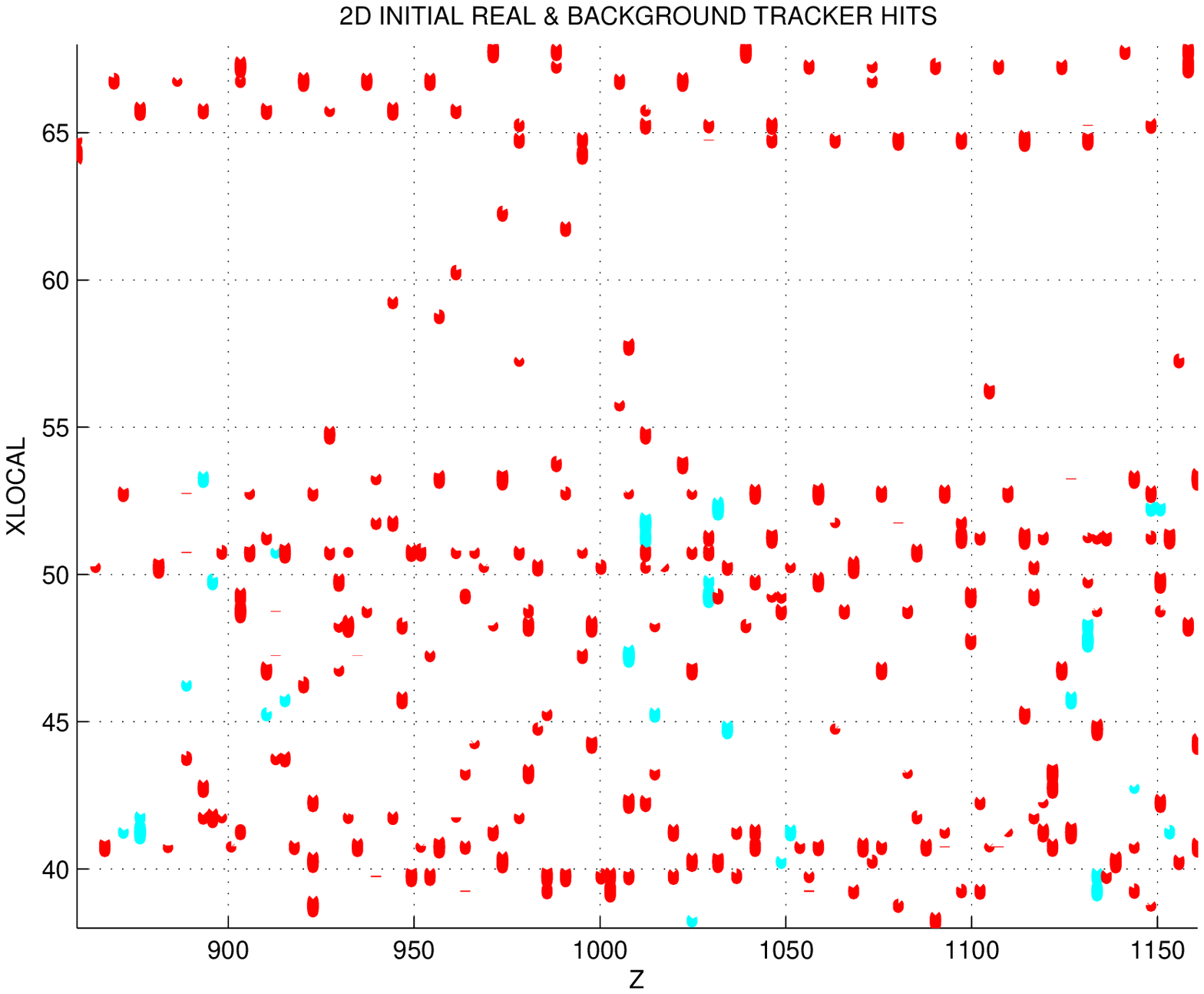}
  \includegraphics[width=0.5\textwidth]{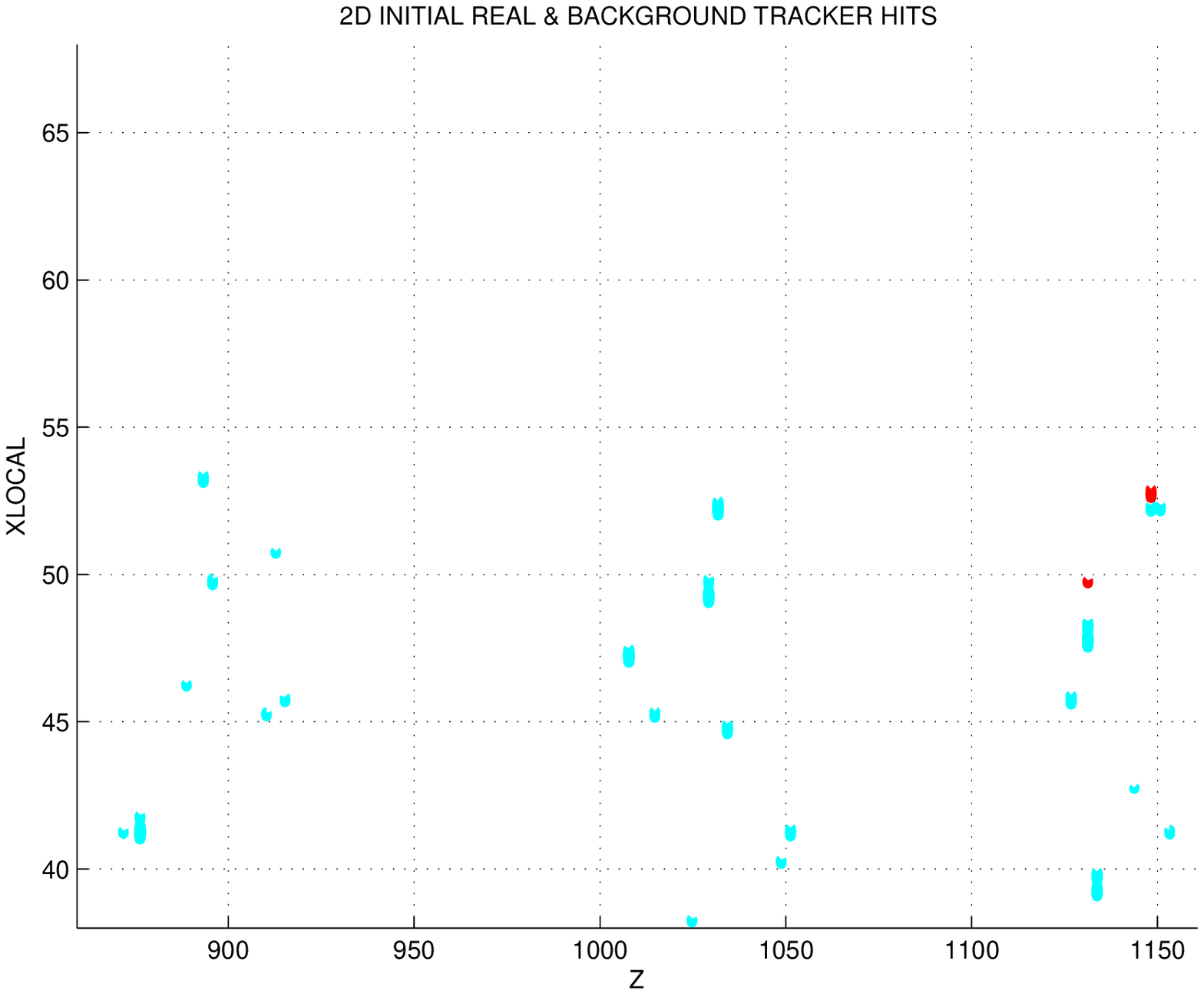} }}
  \caption[Short caption.]{
 Plot of signal (blue dots)  and  background (red dots) tracker 
 hits before (left) and after the pattern 
 recognition procedure without drift time (right).
}
\label{fig:sel_5}
\end{figure}

Figure  \ref{fig:sel_5} shows the multiple view superposition  of tracker hits for the sample event
with 29 signal and 260 background hits before and after the pattern recognition procedure.
There are no missed signal hits and two surviving background hits
in this case.

\subsection{Kalman filter application to track reconstruction}

The Kalman filter (KF) \cite{rkalm,kalman}  addresses the general problem of trying to
estimate at different points ($1\leq k\leq n$) the state
$\bf{x}_{k}$ of a discrete process that is governed by the
linear stochastic difference equation

\begin{equation}
\bf{x}_{k}=\bf{F}_{k-1}\bf{x}_{k-1}+\bf{w}_{k-1}
\label{1ap}
\end{equation}

with a measurement $\bf{m}_{k}=\bf{H_{k}}\bf{x}_{k}+\bf{\varepsilon} _{k}$. 
In the system of equations (\ref{1ap}) 
stochastic processes such as multiple
scattering and bremsstrahlung are taken into
account by the process noise $\bf{w}_{k}$.
The measurement noise is represented by  $\bf{\varepsilon} _{k}$.
 $\bf{Q}_{k}$ and $\bf{V}_{k}$  are process noise and
measurement noise covariances, respectively.
In the absence of the last term Eq.(\ref{1ap}) is the standard
equation of motion with a propagator $\bf{F}_{k-1}$ (transport
matrix). 
The $\bf{F}$ matrix propagates the state vector on
one measurement plane to the state vector on the next plane combining position
information with directional information.

For a particle moving in a uniform magnetic field (see Fig. \ref{fig:setup}) 
one has to choose the state vector
parameters, define the initial state vector and calculate the
transport matrix $\bf{F}$, the projection matrix $\bf{H}$, and
the noise matrix $\bf{Q}$.
The state vector can be chosen
in the form ${\bf{x}_{k}}=(x,y,t_{x},t_{y},1/p_{L})$ where $x, y$
are the track coordinates in the tracker system, and
$t_{x}=p_{x}/p_{L}$, $t_{y}=p_{y}/p_{L}$ define the track
direction.
   The projection matrix  is given by ${\bf{H}}=(\cos\alpha ,\sin \alpha ,0,0,0)$,
where $\alpha$ is a tracker view angle.

   Due to multiple scattering the absolute value of electron momentum
remains unaffected, while the direction is changed. This
deflection can be described using two orthogonal scattering
angles, which are also orthogonal to the particle momentum \cite{mank}.
 In terms of these variables the noise
matrix is given by
\vspace{-0.3in}
$$
\bf{Q}_{k}=<\Theta ^{2}>(t_{x}^{2}+t_{y}^{2}+1) \left.\left(
\begin{array}{ccccc}
0 & 0 & 0 & 0 & 0 \\
0 & 0 & 0 & 0 & 0 \\
0 & 0 & t_{x}^{2}+1 & t_{x}t_{y} & t_{x}/p_{L} \\
0 & 0 & t_{x}t_{y} & t_{y}^{2}+1 & t_{y}/p_{L} \\
0 & 0 & t_{x}/p_{L} & t_{y}/p_{L} & \frac{(t_{x}^{2}+t_{y}^{2})}{
p_{L}^{2}(t_{x}^{2}+t_{y}^{2}+1)}
\end{array}
\right.\right)$$

   For the variance of the multiple scattering angle the well-known
expression is used

\begin{equation}
\nonumber
<\Theta^{2}>=(p_{0}/p)^{2}[1+0.038\ln(t/X_{R})]t/X_{R}
\end{equation}

where $p_0 =$ 13.6 MeV, $X_{R}$ is a radiation length and  t is a distance traveled by
the particle inside a scatterer.
   Energy losses are taken into account by

\begin{equation}
\nonumber
p\prime = p - <dE/dx>t.
\end{equation}

There are three types of operations to be performed in the analysis of a track.
Prediction $\bf{x}_{k}^{k-1}$ is the estimation of the ``future" state
vector at
position $``k"$ using all the ``past" measurements up to and including $``k-1"$.

Filtering $\bf{x}_{k}^{k}$ is the estimation of the state vector at
position $``k"$ based upon all  ``past" and ``present" measurements
up to and including $``k"$.
At each step the filtered $\chi ^{2}$ is calculated.
The total $\chi ^{2}$ of the track is given by the sum of the $%
\chi _{k}^{2}$ contributions for each plane.

Smoothing $\bf{x}_{k}^{n}$ is the estimation of the ``past" state
vector at
position $``k"$ based on all $``n"$ measurements taken up to the present time.
The filter runs
backward in time updating all filtered state vectors on the basis
of information from all n planes. The mathematical equations describing
these operations are given in \cite{rkalm,kalman}.

The prediction and filtering are applied consecutively  to all points.
When the last point is reached the smoothing is applied to all previous points.
The result of smoothing  is a reconstruction of the state vector 
$\bf{x}_{k}^{n}$ which defines particle coordinates and momentum in each plane.
The KF approach described above works for a single point in each plane. Background hits create
multiple competitive point in each plane. In this case a different approach based on the KF 
will be applied as described below. 

\subsection{Pattern recognition with drift time }

The second stage in the pattern recognition procedure uses the
hits selected and 
fitted helix obtained in the first stage 
and adds drift time information for  the selected hits.
The pattern recognition procedure can be improved by taking
into account the measured drift time $t_{i}^{meas}$ which is related to
the radial distance r at the closest approach to the straw wire.
The errors ($\sigma $) in radius measurements
are taken to be 0.2 mm. This radius r carries an ambiguity as to
whether the track passed left or right of the wire. 
Left and right points are extracted 
from the intersections of a normal to the
helix through the straw center and the circle of the radius r. 

A deterministic annealing filter (DAF) \cite{daf, rev} 
is applied to suppress background hits further.
The DAF is a Kalman filter  with re-weighted observations.
The propagation part of DAF is
identical to the standard Kalman filter. 
In addition to background hits a left-right ambiguity for given hit creates multiple
competing points at each KF step. 
At the DAF step all competing points are assigned to a single layer with  weights.
The filtered estimate (measurement update)
$\bf{x}_{k}^{k}$ at layer k is calculated as a weighted mean of the
prediction $\bf{x}_{k}^{k-1}$ and the observations 
{$\bf m_{k}^{i}, i=1,2,...n_{k}$}.
  
By taking a weighted mean of the filtered states  
at every layer a prediction for the state vector 
$\bf x_{k}^{n*}$ along with its covariance matrix $\bf {C}_{k}^{n*}$ 
is obtained, using all hits except the ones at layer k. 
Initially all assignment probabilities for the
hits in each layer 
are set
to be equal but based on the estimated state vector 
$\bf x_{k}^{n*}$ and its covariance 
matrix, the assignment probabilities of all competing hits are then 
recalculated in the following way:

\begin{equation}
\nonumber
{\bf {p}_{k}^{i}} \sim \varphi(\bf{m}_{k}^{i};
\bf{H}_{k}\bf x_{k}^{n*}, \bf{V}_{k}+\bf{H}_{k}\bf{C}_{k}^{n*}\bf{H}_{k}^{T})
\label{4bp}
\end{equation}

where $\varphi$ is a multivariate Gaussian 
probability density, $\bf{V}_{k}$ is the variance of the observations.
If the probability falls below a certain threshold, the hit is 
considered as background and is excluded from the list of the
hits assigned to the track. 
At the initial step we cannot be sure of calculated probabilities
due to insufficient information for
the filter. This problem is overcome by adopting a simulated annealing     
iterative procedure \cite{daf, rev}  which allows
avoiding a local minimum and finding the global one corresponding to
the minimum chi-square for the track.
The annealing schedule is chosen for the iteration n in the form
${\bf V_{n}}={\bf V} (1 + \frac{C}{f^{n}})$ where the annealing factor 
$f > 1$ and constant $C \gg 1$. 
This insures that the initial 
variance  is well above the nominal value 
${\bf V}$ of the observation error but the final one tends 
to ${\bf V}$. After each iteration the assigned 
probabilities exceeding the threshold are normalized to 1
and used again as weights in the next iteration, and so on.
The iterations generally are stopped if the relative change in 
chi-square is less than a corresponding control parameter
(typically of the order 0.01). 
Since we are dealing with a stochastic process, the best result
can be reached repeating the DAF procedure for a few different
annealing factors f (1.4 and 2) and then choosing the result 
corresponding to the minimum chi-square. 

The application of the DAF procedure mainly allows  resolving the left-right
ambiguity. The DAF chooses wrong left-right points for 6$\%$ of the tracker hits
if the radial distance r at the closest approach to the straw wire is greater then 0.25 mm.
This corresponds to an average two tracker hits with the wrong left-right assignment per event. 
The number of
background hits remaining after the second stage averages 0.38 hits per event in comparison with the
primary 300 hits on average. The final background suppression factor is thus
300/0.38 $\approx$ 800.
Some of the signal hits are lost due to the selection
(0.8 hits or 2.7$\%$ per event on average).
The momentum resolution of the tracker after application of the DAF procedure is 0.25 MeV
for the $\mu \to e$ conversion process.
The  pattern recognition procedure described above is thus effective in the rejection
of background hits and
also provides a good starting point for the track reconstruction.
However,  further improvement in the left-right assignment is necessary to achieve 
better resolution.  
For this we use the procedure described in the following section.

\section{Momentum reconstruction for the $\mu \to e$ conversion process}

The momentum reconstruction is based on the hits selected by the
two stage pattern recognition procedure described above.
For straw drift tubes there is a set of left and right point for each straw hit due to
the left-right ambiguity. 
An  extreme possibility would be 
to apply the Kalman filter to  all possible  $2^N$ 
combinations to reconstruct a track with N hits and to choose the combination with the best $\chi^{2}$. 
This pure combinatorial approach may be the best but it is computationally unfeasible.
The other  extreme could be
to apply the Kalman filter at each step to left and right points 
 and to select the point providing the best $\chi^{2}$.
However this algorithm often selects  wrong points.

It has been proposed in the literature \cite{daf, rev} to use  
a Gaussian sum filter (GSF) 
for solving the assignment problem in a track detector with 
ambiguities \cite{GSF}.
In the GSF approach \cite{Kit}, both process noise and 
observation errors are modeled by a mixture of Gaussian densities
and in the resulting algorithm several
Kalman filters run in parallel.
However the application of GSF as a momentum reconstruction 
procedure to this tracker does not provide sufficiently good 
results because of large gaps in the measured trajectory.
In the tracker hits are distributed
approximately uniformly inside lobes in the z-direction, but
there are large gaps about 70 cm between lobes due to
the motion of the  electron outside the sensitive area of the 
tracker. A Kalman filter prediction step corresponding to 
a gap often leads to significant deviations from the true hit.
As a result, in the GSF procedure the true component often is
suppressed so significantly that it can not be recovered by
subsequent Kalman filter steps.
Therefore we need a different approach to treat  the left-right ambiguity
in the tracker.    

\begin{figure}[htb!]
  \centering
  \includegraphics[width=0.7\textwidth]{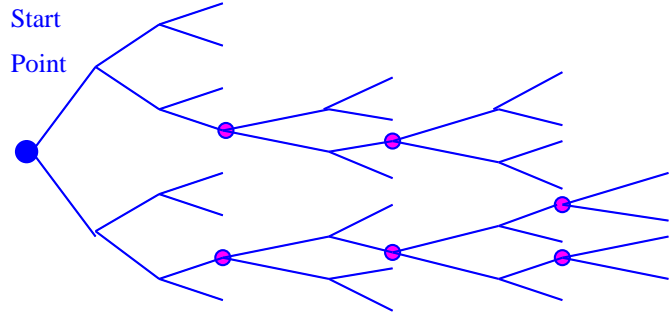} 
  \caption[Short caption.]{
 Graph 2$\rightarrow$8 for CDF procedure.  
}
\label{fig:ccf}
\end{figure}

A combinatorial drop filter (CDF) has been developed 
to improve the momentum reconstruction and efficiency for the tracker.
The CDF algorithm uses forced minimization of
the number of combinations when it reaches some maximum.  
So in the CDF approach the number of combinations always oscillates 
between $N_{min}$  and $N_{max}$.
That means that the Kalman filter runs with paths increasing in number
with the combinatorial growth of combinations,
but each time  the number of combinations reaches $N_{max}$, only the $N_{min}$ 
combinations with the best $\chi^2$ are retained.
The CDF approach is illustrated by  the graph $N_{min}$$\rightarrow$$N_{max}$ 
(2$\rightarrow$8) in Fig. \ref{fig:ccf}.
Each vertex represents either a left or right
point. Components surviving after the drop are indicated by small (magenta)
circles. 
In the actual  strategy 
the number of retained components 
was chosen 
to be 8  and the maximum number of components 
to be 32.

The procedure begins with 
the first eight straw hits and builds $2^8$ (256)
possible hit combinations corresponding to left-right points.
The Kalman filter  forward and backward procedures are applied
to these combinations. Only a few combinations which satisfy a
rather loose $\chi^2$ cut,
are retained. 
This step provides a certain precision for state-vector
components. 
Each retained combination containing 8 points provides a starting point for
the CDF.

\begin{figure}[htb!]
  \centering
  \includegraphics[width=0.6\textwidth]{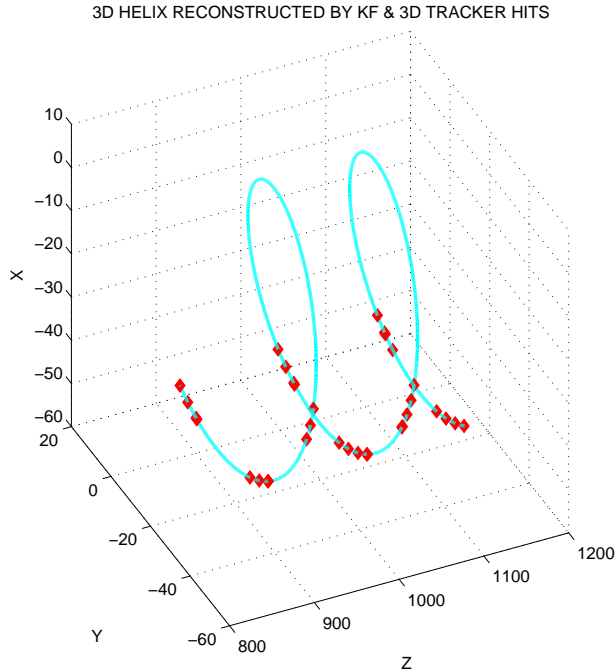} 
  \caption[Short caption.]{
 The 3D trajectory reconstructed for a sample event.
}
\label{fig:fig_3D1}
\end{figure}

The Kalman filter reconstructs a trajectory of a particle
in three dimensions. The trajectory is bent
each time it crosses a tracker plane due to multiple 
scattering. 
Therefore, the reconstructed track is a set of
helices that intersect at the  planes.
Fig. \ref{fig:fig_3D1}
shows the 3D trajectory  reconstructed for a
sample event.

\begin{figure}[htb!]
  \centering{\hbox{
  \includegraphics[width=0.5\textwidth]{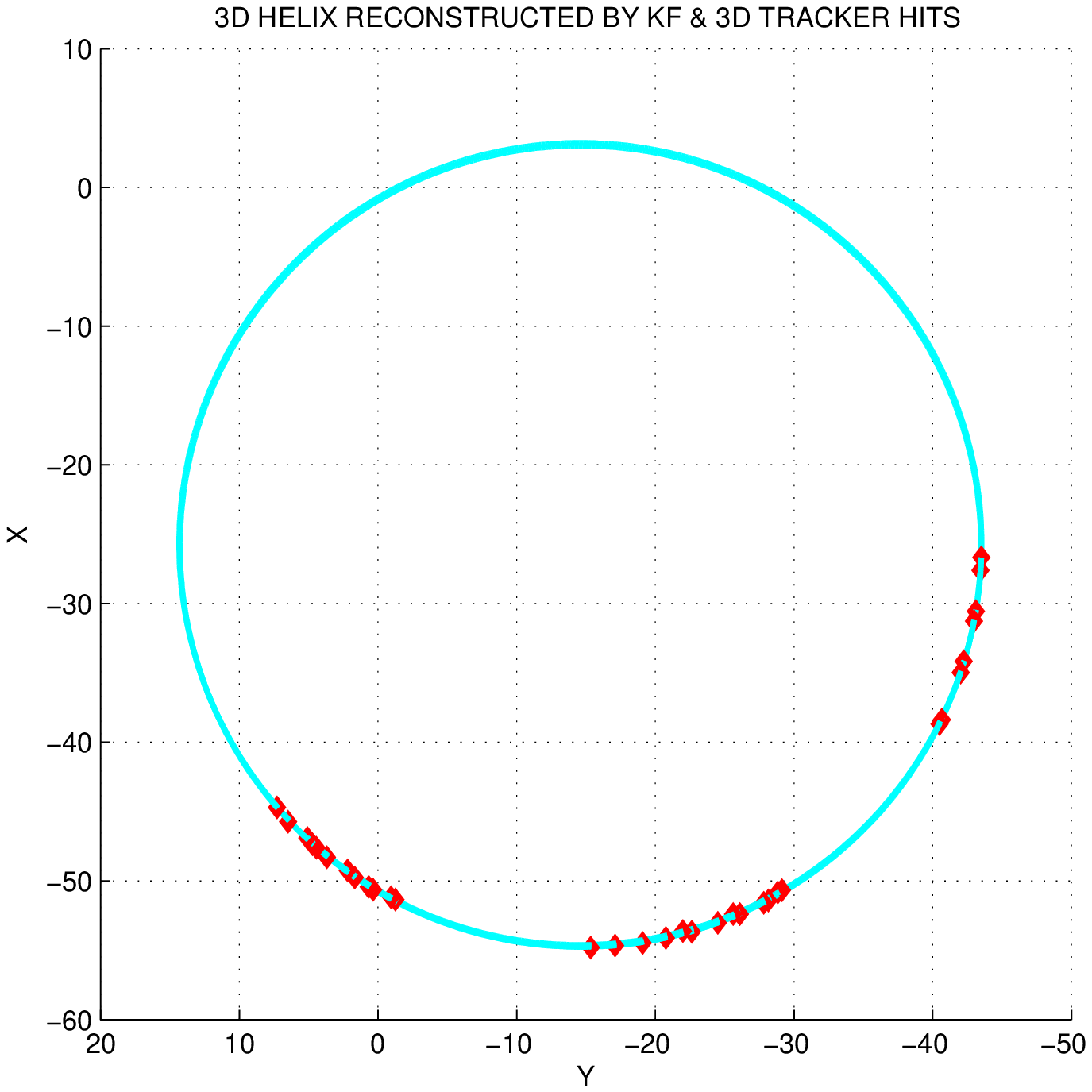}
  \includegraphics[width=0.5\textwidth]{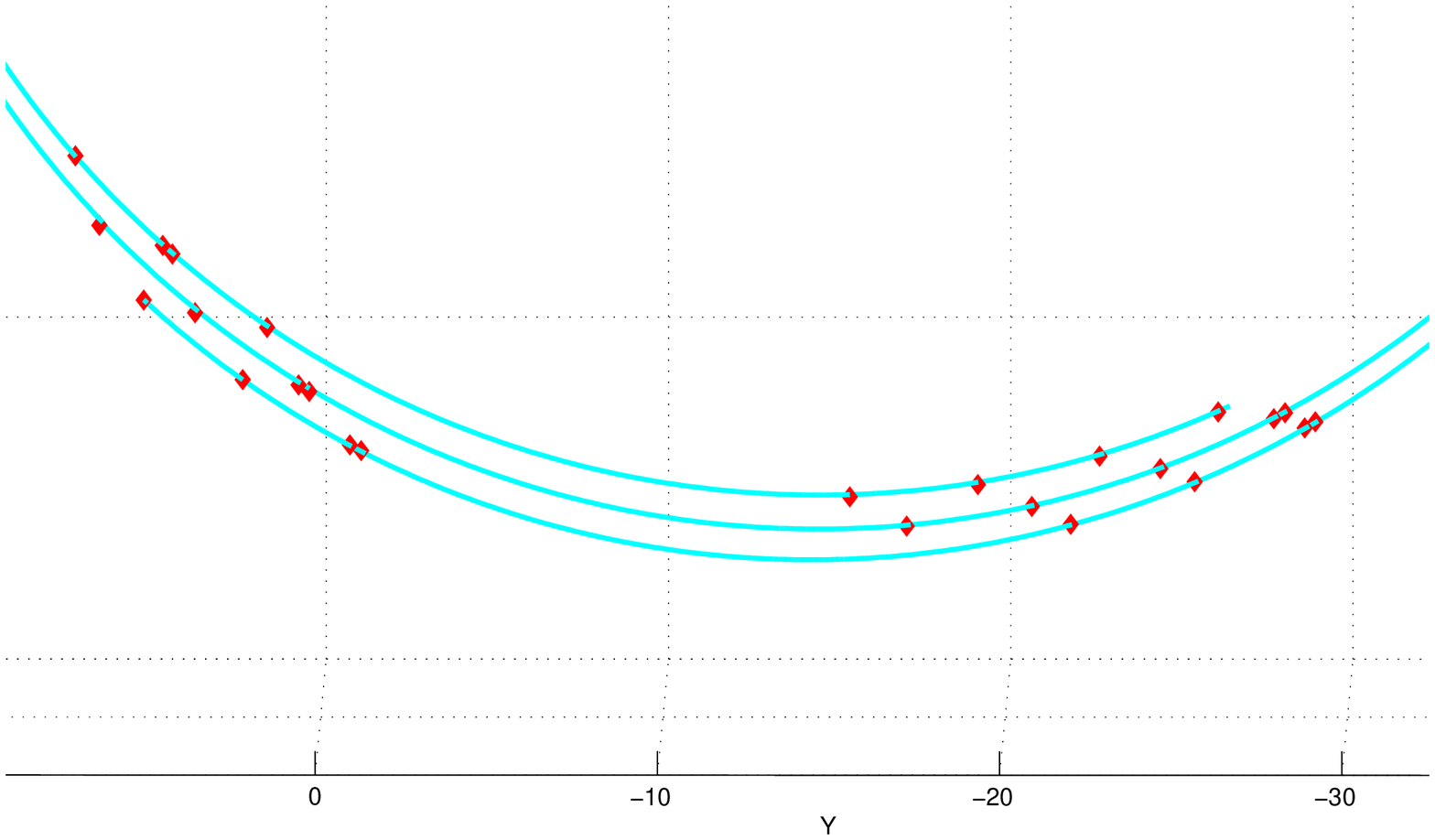} }}
  \caption[Short caption.]{
The transverse projection of the 3D trajectory reconstructed for a sample event (left).
Enlargement of the bottom part of the trajectory (right).
}
\label{fig:fig_3D3}
\end{figure}

On this scale the trajectory looks  like a helix, but it
consists of many helix parts.
Fig. \ref{fig:fig_3D3} (left)
shows the transverse projection of the 
trajectory for the sample event. 
In this projection the trajectory looks approximately 
like a circle.
However if a specific region of Fig. \ref{fig:fig_3D3} (left)  is magnified 
one can see in Fig. \ref{fig:fig_3D3} (right)
that the shape of the
circle is distorted due to multiple scattering and energy
loss.  Three arcs of the trajectory are clearly 
seen in the Figure.

There are two main tracker reconstruction features:
momentum and angle resolutions.
 The momentum resolution is defined from
the distribution of the difference between simulated and reconstructed momentum.
 The angle resolution is defined from
the distribution of the angle between simulated and reconstructed  momentum.

\begin{figure}[htb!]
  \centering{\hbox{
  \includegraphics[width=0.5\textwidth]{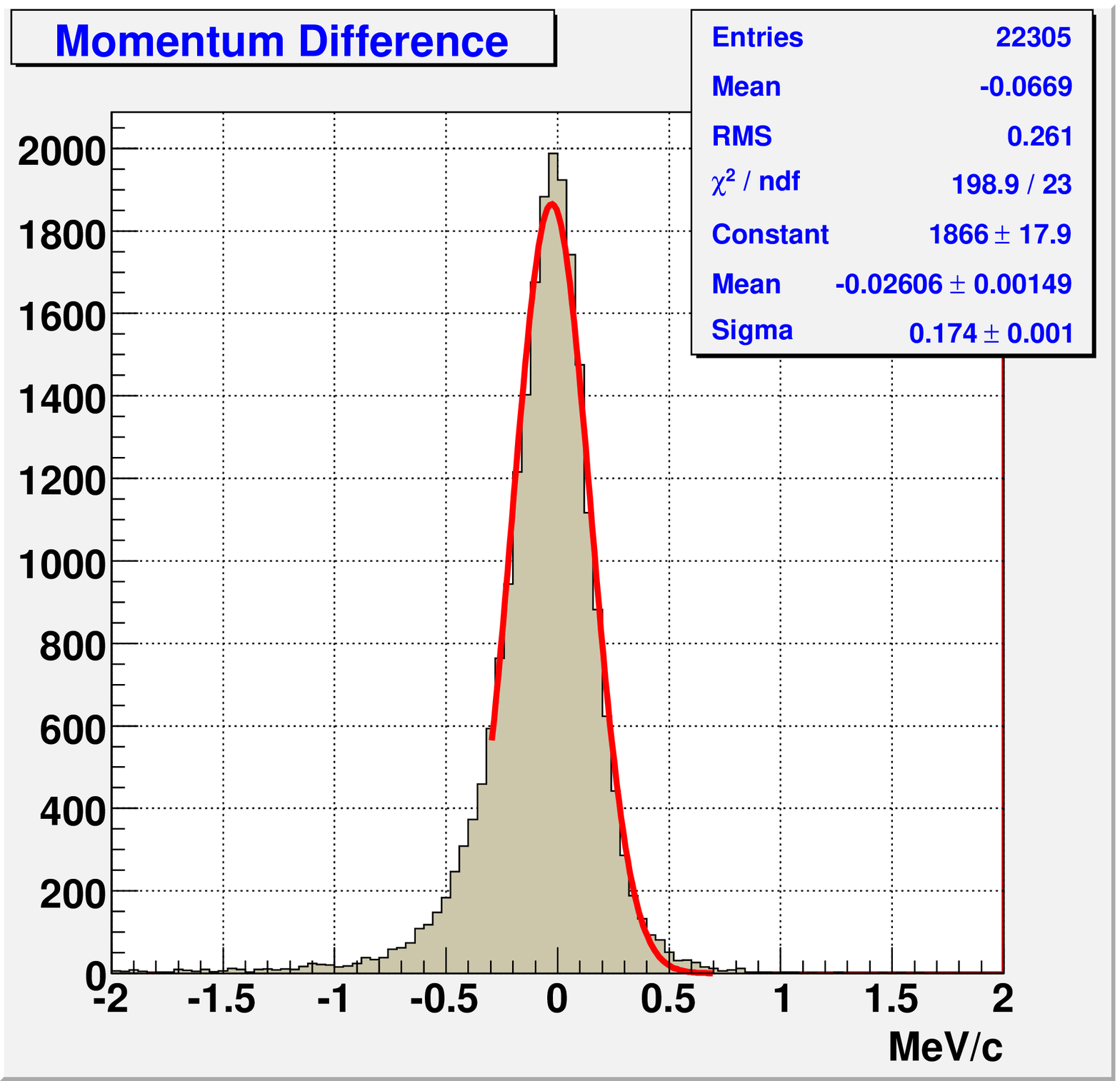}
  \includegraphics[width=0.5\textwidth]{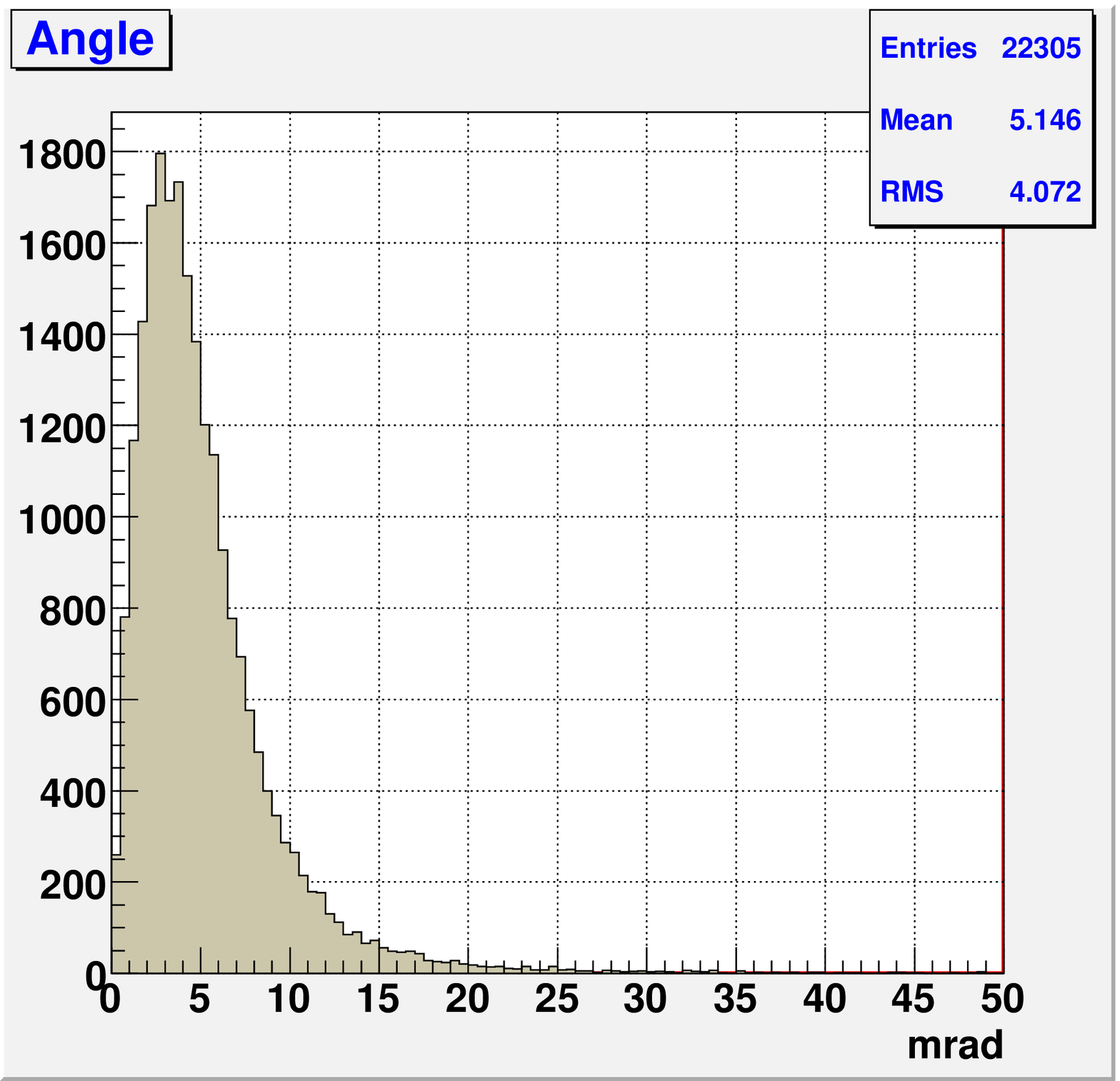} }}
  \caption[Short caption.]{
 Reconstruction for $\mu \to e$ conversion process.
 The distribution of the difference between the reconstructed 
 and simulated   momentum (left). The distribution of the angle  between the reconstructed 
and simulated   momentum (right).
}
\label{fig:mue}
\end{figure}

In this study $10^5$ events were simulated and reconstructed.
The CDF procedure chooses wrong left-right points for 4$\%$ of the tracker hits
(to be compared with 6$\%$ for the DAF)
if the radial distance r at the closest approach to the straw wire is greater then 0.25 mm.
The distribution of the difference between the initial momentum  
reconstructed by
the Kalman filter and the simulated initial momentum is shown in
Fig. \ref{fig:mue} (left) for the $\mu \to e$  conversion process. 
According to this distribution the intrinsic tracker resolution is
$\sigma$ = 0.17 MeV if one fits the distribution by a Gaussian. 
Note that this tracker resolution is by about 30$\%$ better than the
resolution provided by the DAF procedure.

 The distribution of the angle  between the reconstructed 
and simulated   momenta is shown in Fig. \ref{fig:mue} (right).
The  most  probable value of the angle distribution is 3 mrad.
A calculation gives an average angle of multiple scattering in a straw 
to be 1.8 mrad for the $\mu \to e$  conversion process.
This is consistent with the expectation that the angular resolution is
defined by multiple scattering in a single tracker straw.

\subsection{Momentum reconstruction for the $\mu^+ \to e^+ + \gamma$ process }

\begin{figure}[htb!]
  \centering{\hbox{
  \includegraphics[width=0.5\textwidth]{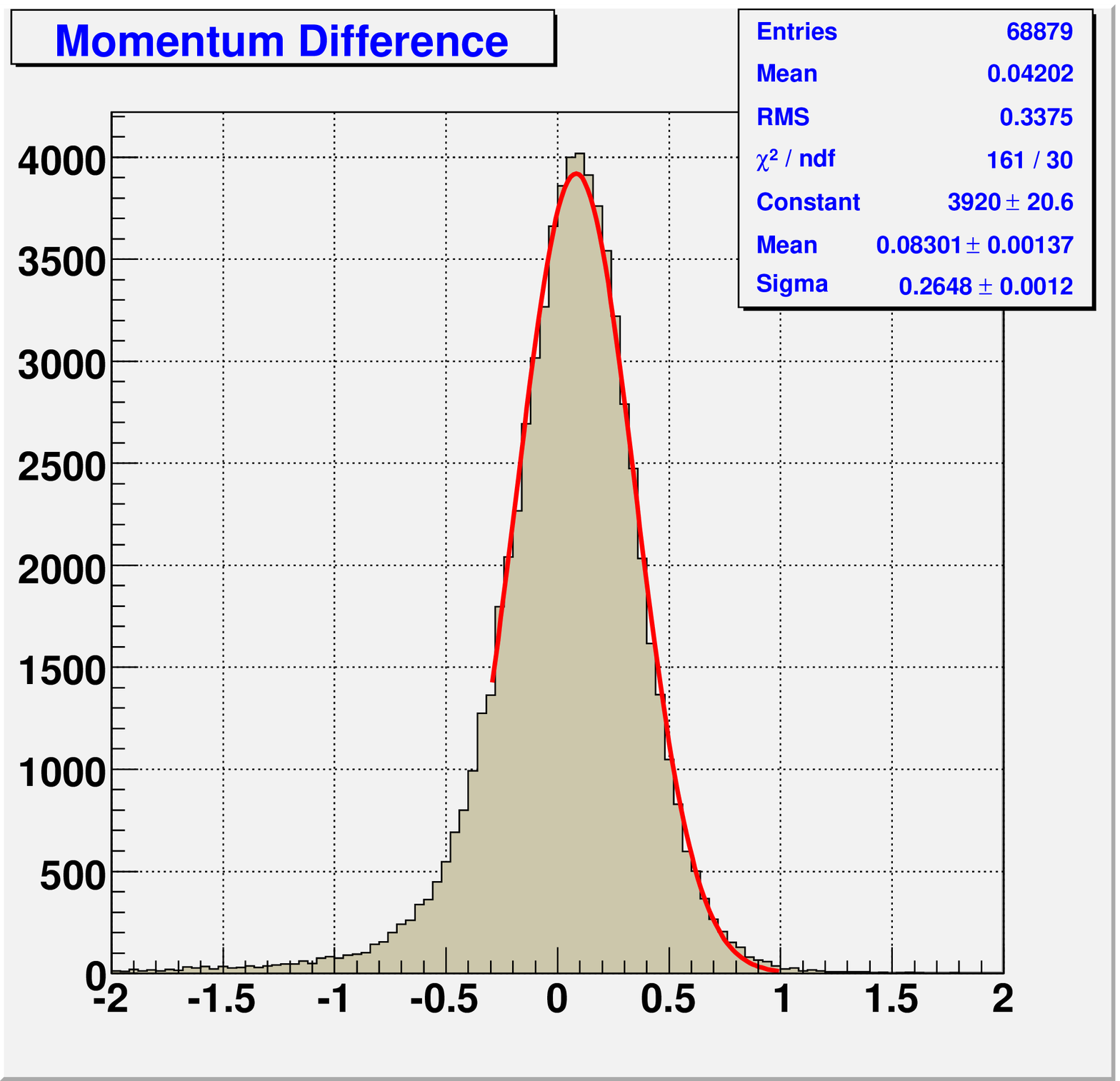}
  \includegraphics[width=0.5\textwidth]{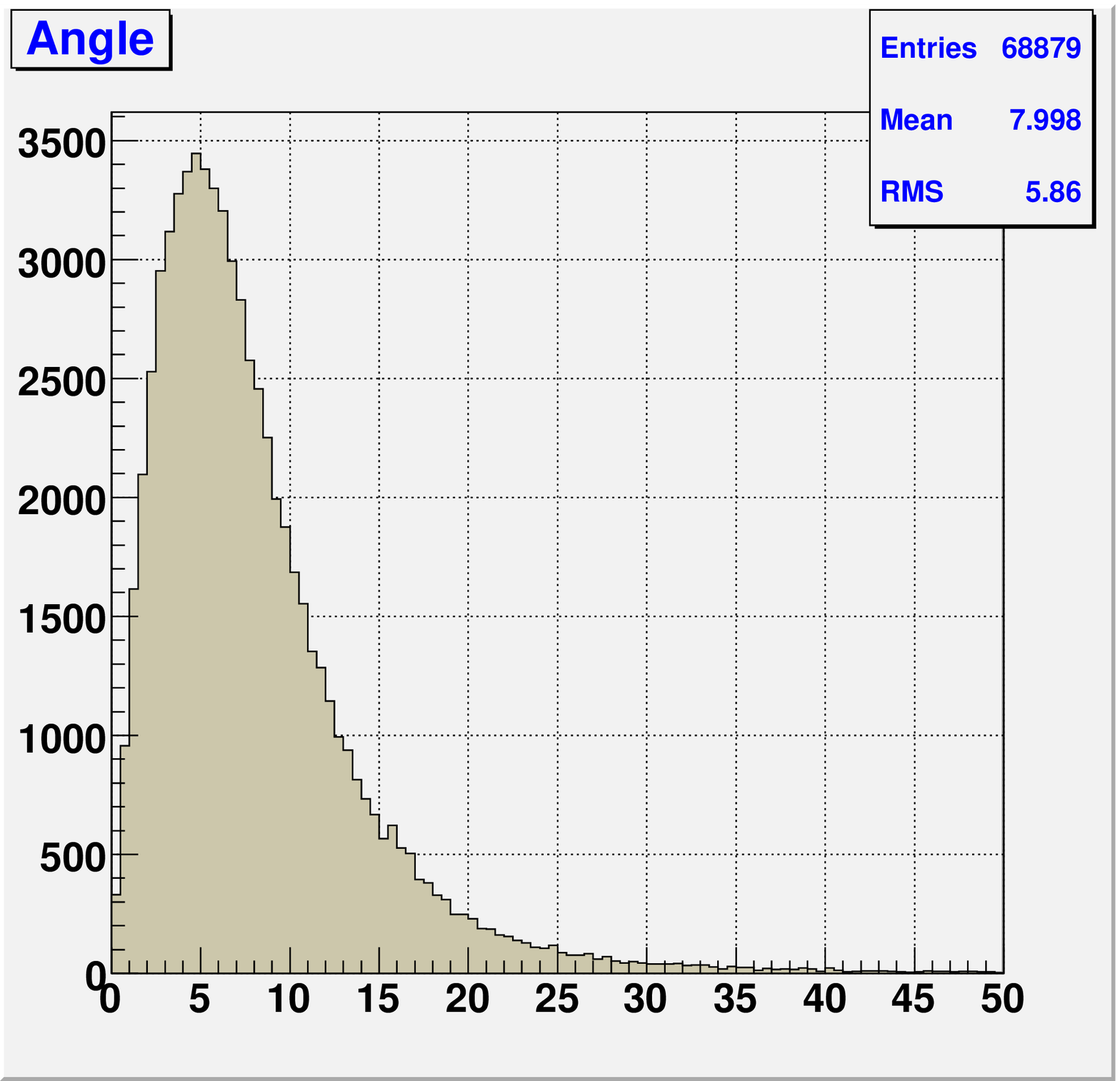} }}
  \caption[Short caption.]{
 Reconstruction for the $\mu^+ \to e^+ + \gamma$ process:
 The distribution of the difference between the reconstructed 
 and simulated   momentum (left). The distribution of the angle  between the reconstructed 
and simulated   momentum (right).
}
\label{fig:meg}
\end{figure}

As mentioned in the introduction, this study and the one in the next section are less complete 
than that carried out for the $\mu \to e$ conversion process. No background hits were added to simulated events.
The second tracker layout and field configuration were used for the momentum reconstruction study 
of the  $\mu^+ \to e^+ + \gamma$ process without background.   
The tracker is placed in a magnetic field of 0.5T which is half that  
used for  the  $\mu \to e$ conversion setup.
The  $\mu \to e$ conversion and  the 
$\mu^+ \to e^+ + \gamma$ processes have   monochromatic charged leptons in the final state
whose energies  differ by a factor of 2.
Due to the twofold change in the magnetic field the charged lepton trajectories for the 
$\mu^+ \to e^+ + \gamma$
and the $\mu \to e$ conversion  processes are similar.
The pattern recognition and momentum reconstruction procedures described above 
were applied to positrons of the $\mu^+ \to e^+ + \gamma$ process.
 The distribution of the difference between the initial momentum  
reconstructed by
 the Kalman filter and the simulated initial momentum  is shown in
Fig. \ref{fig:meg} (left). 
Fitting this distribution to a Gaussian gives an intrinsic tracker resolution of
$\sigma$ = 0.26 MeV. 
 The distribution of the angle  between the reconstructed 
and simulated   momentum is shown in Fig. \ref{fig:meg} (right).
The  most  probable value of the angle distribution is 5 mrad.
A calculation gives an average angle of multiple scattering in a straw to be 3.6 mrad 
for a positron in the $\mu^+ \to e^+ + \gamma$ process.

\subsection{Momentum reconstruction for the $\mu \to 3e$ process }

The same setup as for the $\mu^+ \to e^+ + \gamma$ process 
was used for a momentum reconstruction study of the $\mu^+ \to e^+ + e^+ + e^-$ process.   
This  process has in the final state
an electron and two positrons with  energies from 0 to $m_{\mu}$/2 and  an average energy equal 
to one third of the muon mass.
The pattern recognition and momentum reconstruction procedure applied for $\mu \to e$  conversion 
and $\mu^+ \to e^+ + \gamma$ processes was based on a single track search algorithm. 
The same procedure was applied separately for each lepton track of the $\mu^+ \to e^+ + e^+ + e^-$. 
The observable quantity of the process, total scalar momentum of charged leptons
$p_{tot} = \sum {p_i} $, was reconstructed.
 
\begin{figure}[htb!]
  \centering{\hbox{
  \includegraphics[width=0.5\textwidth]{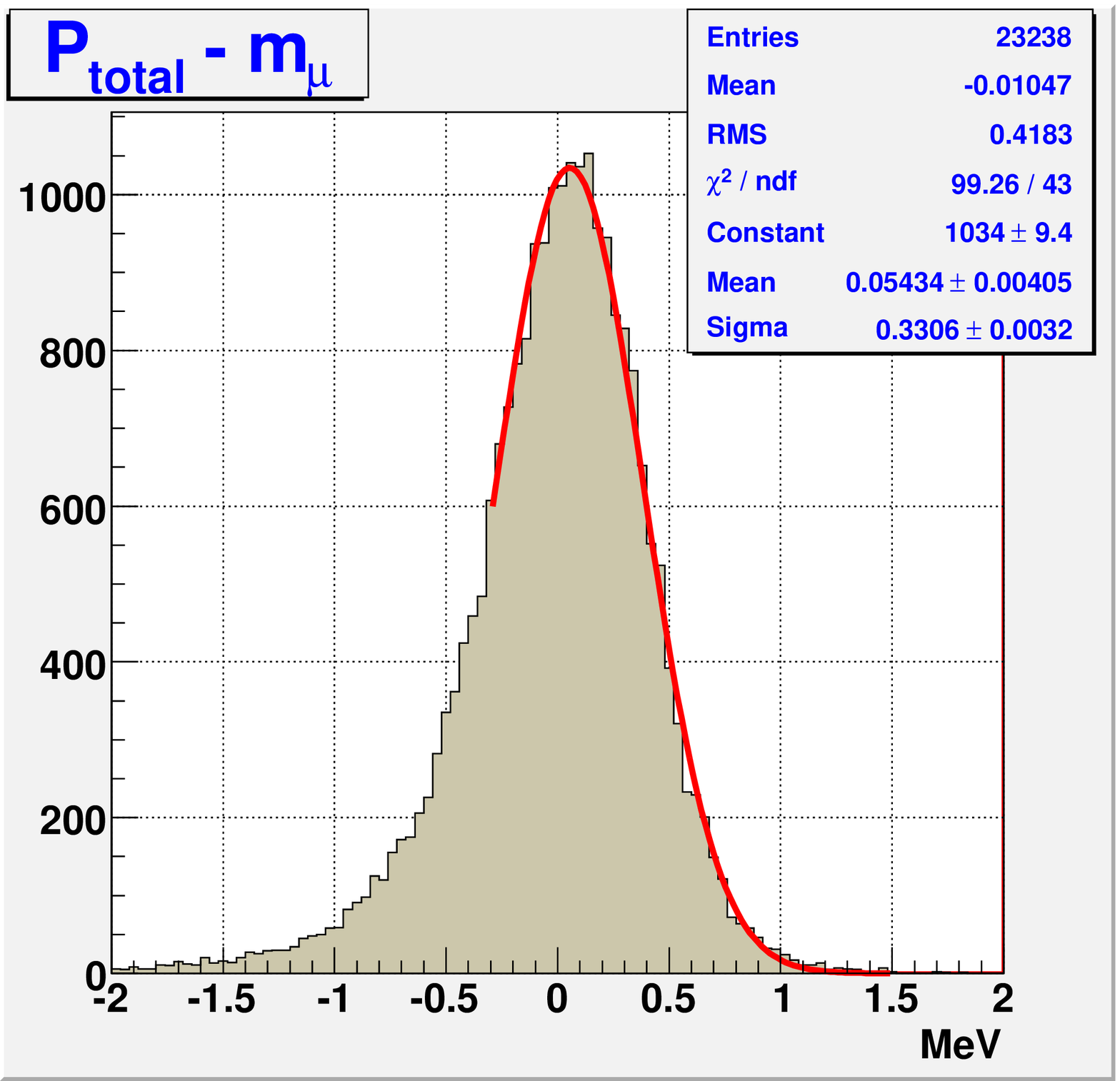}
  \includegraphics[width=0.5\textwidth]{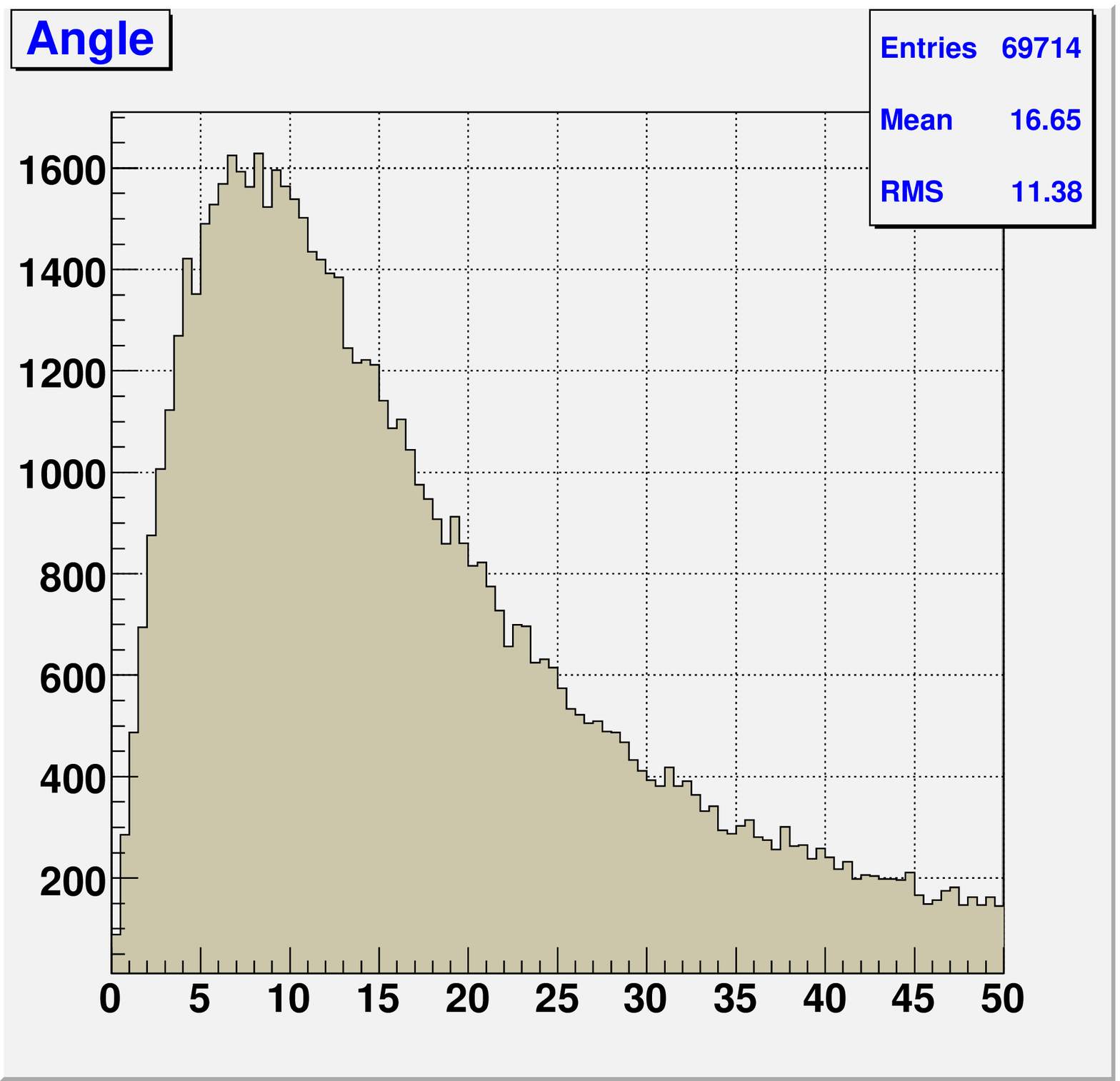} }}
  \caption[Short caption.]{
 Reconstruction for the $\mu \to 3e$ process:
 The distribution of the difference between   
reconstructed total scalar momentum  
and muon mass (left). The distribution of the angle  between the reconstructed 
and simulated   momentum (right).

}
\label{fig:mu3e}
\end{figure}

 The distribution of the difference between   
reconstructed total scalar momentum  
and muon mass   is shown in
Fig. \ref{fig:mu3e} (left).
Fitting this distribution by a Gaussian gives a
tracker resolution for total scalar momentum  of
$\sigma$ = 0.33 MeV. 
 The distribution of the angle  between the  reconstructed   lepton momentum
and the simulated  momentum is shown in Fig. \ref{fig:mu3e} (right).
The  most  probable value of the angle distribution is 7 mrad.
A calculation gives an average angle of multiple scattering in a straw to be 5.4 mrad 
for charged lepton momentum equals $m_{\mu}$/3.

\section{Conclusion}

A pattern recognition and momentum reconstruction 
procedure based on a Kalman filter technique 
for a straw tube tracker
was developed to 
suppress background and reconstruct track momentum
for  the lepton flavor violating processes $\mu \to e$ conversion,  $\mu^+ \to e^+ + \gamma$ and
$\mu \to 3e$.
The simple modular construction of the tracker allows  meeting requirements for lepton 
momentum measurements for all lepton flavor violation processes mentioned above.

The pattern recognition procedure allows suppressing  the  background  hits
by a factor of 800. Only 3\% of signal hits are lost by this procedure.

The momentum resolution of the tracker to register monochromatic leptons 
was found to be  $\sigma_{p} = $0.17  and 0.26 MeV for the $\mu \to e$ conversion
with background and the $\mu^+ \to e^+ + \gamma$ processes without background, respectively.
The tracker  resolution for the total scalar lepton momentum 
is  $\sigma_{p} = $ 0.33 MeV for the $\mu \to 3e$ process without background.
For  the $\mu \to e$ conversion,  $\mu^+ \to e^+ + \gamma$ and
$\mu \to 3e$
 processes the most  probable values of an
angle between the simulated and reconstructed   momenta (3, 5 and 7 mrad)
are in a good agreement with multiple scattering angles in a single tracker straw. 
The obtained tracker resolutions allow an increase in sensitivity to the branching ratios 
for these processes by
a few orders of magnitude over current experimental limits.

We wish to thank A. Mincer and P.Nemethy
for fruitful discussions and helpful remarks.
This work has been supported by the National Science Foundation under grants PHY 0428662, PHY 0514425
and PHY 0629419.

\newpage

\end{document}